\documentclass{WileyMSP-template}
\usepackage{float}
\usepackage{hyperref}
\usepackage{amsmath}

\begin{document}

\pagestyle{fancy}
\rhead{\includegraphics[width=2.5cm]{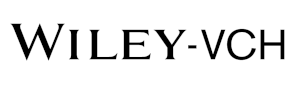}}

\title{Study of Resistive Switching Dynamics and Memory States \\Equilibria in Analog Filamentary Conductive-Metal-Oxide/HfO$_{\rm x}$ \\ReRAM via Compact Modeling}

\maketitle


\author{Matteo Galetta*}
\author{Donato Francesco Falcone}
\author{Victoria Clerico}
\author{Wooseok Choi}
\author{Stephan Menzel}
\author{Antonio La Porta}
\author{Tommaso Stecconi}
\author{Folkert Horst}
\author{Bert Jan Offrein}
\author{Valeria Bragaglia**}

\begin{affiliations}
M. Galetta, D. F. Falcone, V. Clerico, Dr. W. Choi, Dr. A. La Porta, Dr. T. Stecconi, Dr. F. Horst, Prof. B. J. Offrein, and Prof. V. Bragalia \\
Science of Quantum and Information Technology, IBM Research Europe-Zurich, Rüschlikon CH-8803, Switzerland\\ 
E-mail Address: ata@zurich.ibm.com*, vbr@zurich.ibm.com**
\end{affiliations}

\begin{affiliations}
Dr. S. Menzel\\
Peter Grünberg Institute-7, Forschungszentrum Jülich, Jülich DE-52425, Germany
\end{affiliations}



\begin{abstract}

Resistive Random Access Memory (ReRAM) devices offer a promising solution for next-generation non-volatile memory and neuromorphic computing systems. Yet, existing compact models fail to capture analog resistive switching behavior of ReRAM devices. This work presents an advanced physics-based compact model for analog filamentary Conductive-Metal-Oxide (CMO)/HfO$_{\rm x}$ ReRAM, capable of reproducing switching characteristics over a broad range of operating conditions. Compared to the state-of-the-art, the model extends the dynamic interplay between ion migration and electron hopping, while also accounting for parasitic resistive elements. Simulations of various voltage inputs are tested to reproduce quasi-static $I$-$V$ curves, SET switching kinetics under single-pulse programming conditions, and analog accumulative conductance modulation upon bipolar identical pulse streams.
Additional simulations reveal the physical criterion underlying the stabilization of the CMO/HfO$_{\rm x}$-based ReRAM memory state around the equilibrium point, namely symmetry point, under pulsing conditions when a fading memory mechanism emerges. Building upon the evidence of such equilibrium stabilization under pulsing and quasi-static conditions, a procedure is established to visualize and map equilibrium memory states across different input domains.
The physical model supports design optimization of switching behavior for analog neuromorphic systems and non-volatile memory architectures. It also enables accurate integrated circuit simulations with CMO/HfO$_{\rm x}$-based ReRAM technology.

\end{abstract}


\section{Introduction}

In recent years, as Complementary Metal-Oxide-Semiconductor (CMOS) devices continued to scale down, memory units based on standard technologies—such as Static Random Access Memory (SRAM), Dynamic Random Access Memory (DRAM), and flash—have struggled to keep pace with the performance improvements achieved by CMOS processors \cite{epoch2023trendsinmachinelearninghardware}. For advanced technology nodes, embedded memory elements exhibit escalating costs \cite{Wang2024}. Some of them have also shown incompatibility with CMOS manufacturing processes \cite{yu2021compute}, struggling to meet the demand for higher memory densities. Additionally, as Field-Effect Transistor (FET) size shrinks, increased leakage currents hinder the performance of standard memories \cite{roy2003leakage}. \\
Meanwhile, memory modules have encountered latency issues in data movement \cite{epoch2024datamovementbottlenecksscalingpast1e28flop} at the architecture level, highlighting the shortcomings of the von Neumann paradigm in handling the massive workloads of modern data-centric applications, such as Artificial Intelligence (AI) \cite{Sebastian2020}. As high-throughput data transfer and processing are essential, the separation of memory and processing units creates performance and energy bottlenecks when executing Deep Neural Networks (DNNs) on conventional von Neumann architectures.
While compute performance demands continue to grow, AI workloads are executed on large, power-hungry Graphic Processing Unit (GPU)-based systems \cite{epoch2023trendsinmachinelearninghardware}. In this context, in-memory computing has emerged as a potential solution, enabling computations within memory elements, hence significantly improving energy efficiency by eliminating data movement \cite{Ielmini2018}. \\

Resistive Random Access Memory (ReRAM) has emerged as a promising alternative to conventional memory technologies, drawing attention due to low cost and high density integration in the Back-End of Line (BEOL) of two-terminal cell structures with low thermal budget requirements \cite{Wang2024, Burr2017, Gong2022, Falcone2025}. Beyond its growing use in embedded memory arrays \cite{Wang2024, lanza2025growing}, ReRAM is also a strong candidate for in-memory computing hardware, thanks to its non-volatility, low power consumption, and fast read/write performance \cite{Burr2017, Wong2012}. \\

ReRAM technology in crossbar array structures can perform Matrix-Vector Multiplications (MVMs)—the core operations in DNNs—with constant time complexity (O(1)). These cross-point architectures allow for highly parallel computation and offer a significant reduction in latency and energy consumption compared to traditional GPU-based digital accelerators \cite{Gokmen2016}. Leveraging these properties, ReRAM-based Resistive Processing Units (RPUs) enable efficient in-memory computing by physically encoding synaptic weights as conductance values in the ReRAM cells. \\

A conventional ReRAM device cell consists of a Metal/Insulator/Metal (M/I/M) stack. Recent advances in material engineering have resulted in improved switching characteristics of HfO$_{\rm x}$-based ReRAM technology through the integration of a properly engineered Conductive-Metal-Oxide (CMO) layer into the device stack. Filamentary CMO/HfO$_{\rm x}$-based ReRAM devices exhibit analog resistive switching and can therefore encode multi-bit values into High, Intermediate, and Low Resistive States (HRS, IRS, and LRS, respectively) \cite{Stecconi2022, Stecconi2024}. In addition, CMO/HfO$_{\rm x}$-based ReRAM technology has demonstrated remarkable retention times for resistive states programmed with low noise, along with excellent endurance—preserving analog properties over more than 10$^{7}$ programming pulses \cite{Stecconi2024}. \\
In the realm of AI, such improved resistive switching characteristics have been exploited to enable accurate and fast analog training of RPU-based DNNs thanks to dedicated algorithms, namely Tiki-Taka \cite{Gokmen2020, Gokmen2021, Rasch2024Agad}. Particularly, this state-of-the-art analog training algorithm assumes that the ReRAM devices support analog bidirectional accumulative conductance updates, and exhibit a memory state where such updates in opposite directions are equal in magnitude, i.e., the Symmetry Point (SP) \cite{Gokmen2020, kim2019zero}. Recently, hardware-algorithm co-optimization has demonstrated the feasibility of fully on-chip training \cite{Gong2022, Falcone2025, Stecconi2024, ClericoISCAS} and long-term inference capabilities \cite{Falcone2025} with CMO/HfO$_{\rm x}$-based ReRAM technology. \\

To fully harness the potential of ReRAM as an emerging memory technology, the availability of device simulation models is essential for its advancements and integration into Integrated Circuits (ICs). Physical models provide critical insights into the microscopic mechanisms behind resistive switching and transport phenomena \cite{Larentis2012, Funck2021, NHFalcone2024}, driving the optimization of electro-structural layer parameters for further technological advancements \cite{IMWFalcone2023, YandongLuo2016}. On the other hand, compact models are vital for describing device behavior at the system level, enabling the design of ICs incorporating ReRAM devices. To meet both demands, physics-based compact models offer an effective balance between accuracy and simulation complexity, supporting broader operating ranges while remaining suitable for IC design. \\

Although various ReRAM device compact models have been proposed in literature, such as the JART model \cite{LaTorre2019} or the Stanford-PKU model \cite{jiang2016compact}, they are tailored for binary switching in metal-oxide ReRAM involving a filament gap formation. Other models targeting analog ReRAM have also been introduced \cite{Liao2020}, based on the assumption of multiple weak filaments within a metal-oxide layer. In contrast, resistive switching in the analog filamentary CMO/HfO$_{\rm x}$ ReRAM relies on a gradual modulation of defect concentration in the CMO layer, over a rigid conductive filament, imposing a need for customized compact models that differ from those in existing literature \cite{LaTorre2019, jiang2016compact, Liao2020}. \\

\begin{figure}[H]
\centering
  \includegraphics[width=\linewidth]{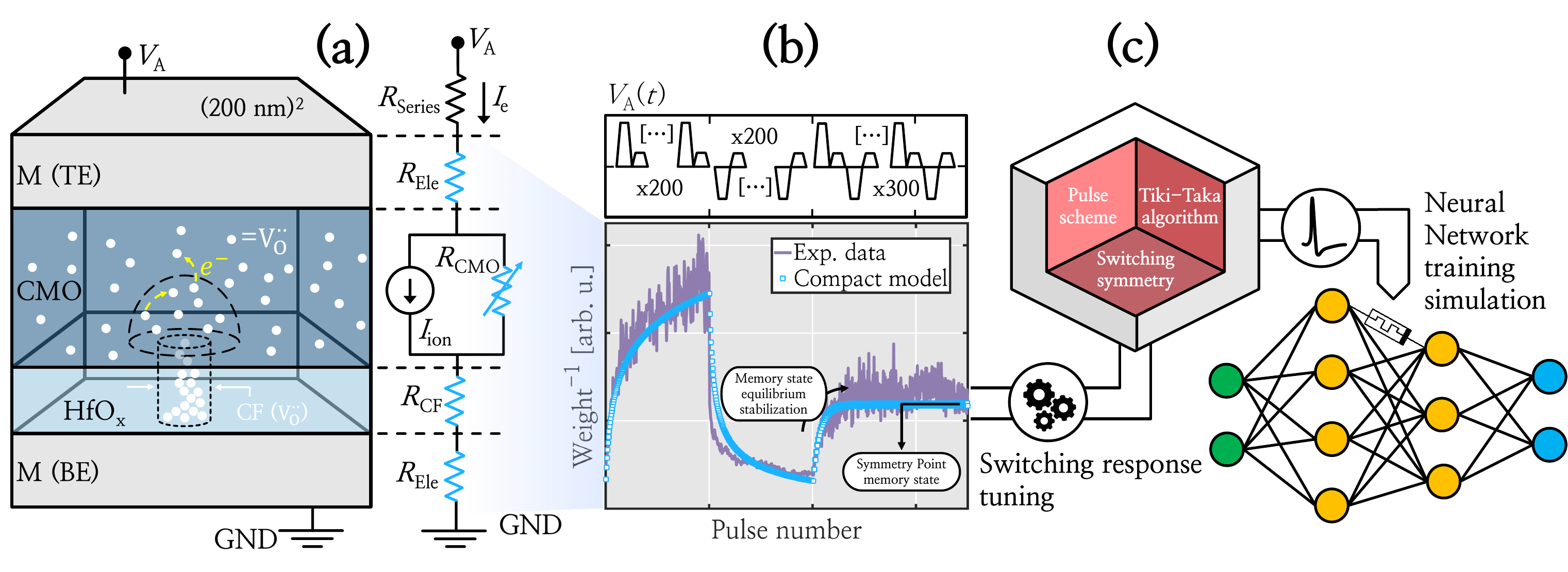}
  \caption{(a) Schematic illustration of the CMO/HfO$_{\rm x}$-based ReRAM structure and the equivalent circuit of the device's compact model used in this work. (b) Analog switching characteristics and stabilization around the symmetry point of the CMO/HfO$_{\rm x}$-based ReRAM device resistance are reproduced through simulations of the experimental pulse scheme. The physical origin of ReRAM device's symmetry point and its dependence on the input pulse scheme are leveraged to study the impact of the switching symmetry in the analog bidirectional accumulative characteristic on MNIST training simulations by using the Tiki-Taka algorithm (c).}
  \label{F1}
\end{figure}

In this work, a compact model tailored for CMO/HfO$_{\rm x}$-based ReRAM is employed to simulate the device characteristics. 3D finite-element models of the same ReRAM device serve as groundwork for the physical assumptions about the electrothermal configuration within the device stack \cite{NHFalcone2024, IMWFalcone2023}. The electron conduction equations used here build upon the analytical description of electronic transport in the non-switching regime, as detailed in \cite{NHFalcone2024}. 
The compact model proposed in \cite{ESSDERC24_Galetta} incorporates thermal and ion migration dynamics to extend the aforementioned analytical framework, enabling a description of the full $I$-$V$ sweep characteristic in a quasi-static voltage input domain. \\
An advanced version of that compact model is presented in this study, integrating key improvements to accurately capture analog resistive switching phenomena in CMO/HfO$_{\rm x}$-based ReRAM. It accounts for the effects of parasitic resistive elements external to the ReRAM cell, corrects ion migration dynamics to prevent unphysical concentrations of oxygen-related defects, and introduces evolving physical parameters that regulate hopping conduction based on the interplay with defect density. A variability component of electronic transport is also included to reflect stochastic conduction non-idealities. In this work, the device model—whose equivalent two-terminal electrical circuit is illustrated in \textbf{Figure \ref{F1}}a—is applied to various electrical input conditions, not addressed in previous studies, to validate its applicability across a broad range of operating time scales. Experimental data of quasi-static $I$-$V$ characteristics, single-pulse programming time measurements, and analog bidirectional accumulative conductance responses are used to test the simulated device behavior.
Further investigation explains the origin of conductance stabilization around an equilibrium point, i.e., the SP (see Figure \ref{F1}b). Inspired by similar approaches \cite{Ascoli2022, Ascoli2024}, the equilibrium point is identified from resistive switching transitions under quasi-static response. This work extends the analysis by studying how the equilibrium stabilization consistently manifests under both pulsed and quasi-static responses. \\
Finally, the simulated switching responses upon different pulse schemes (see Figure \ref{F1}b) are tested for analog Neural Network (NN) training within IBM's AI analog hardware acceleration kit (’aihwkit’) \cite{aihwkit}. This study evaluates the impact of pulse-driven switching symmetry tuning on the learning performance (illustrated in Figure \ref{F1}b and Figure \ref{F1}c) when using the earliest Tiki-Taka variant \cite{Gokmen2020}.

\section{Compact Modeling of Analog Resistive Switching Dynamics}\label{CM}

Former studies \cite{NHFalcone2024, ESSDERC24_Galetta} reported the physical mechanisms behind resistive switching phenomena in CMO/HfO$_{\rm x}$-based ReRAM devices. Such mechanisms rely on a redistribution of oxygen-related defects, where the active layer that undergoes a reversible and non-volatile resistivity change (or memory effect) is the CMO. Under the assumption of sufficiently low switching voltages to avoid interlayer ion migration, the conductive properties of the HfO$_{\rm x}$ layer change only during the electro-forming step. In CMO/HfO$_{\rm x}$-based ReRAM, the electro-forming step involves a relatively large applied bias that induces the formation of a metallic-like nanoscale conductive filament of a highly oxygen-deficient oxide \cite{NHFalcone2024}—schematically illustrated in Figure \ref{F1}a as CF—electrically bridging the bottom electrode and the CMO \cite{Lahkar2025} through a soft dielectric breakdown. Evidence of electronic conduction through the conductive filament, namely filamentary conduction, has been demonstrated in \cite{Falcone2025, Stecconi2022}. In this modeling study, the conductive filament is considered as formed in the HfO$_{\rm x}$ layer with cylindrical shape as a conceptual simplification \cite{Larentis2012}. Electrothermal simulations based on finite element modeling showed that the current is constrained to pass through the ohmic-like conductive filament. Consequently, the presence of a CMO forces the spreading of the electric field from a narrow cylindrical region towards the top electrode \cite{NHFalcone2024}. Accordingly, when a bias is applied, most of the electrostatic field and temperature enhancement experience a confinement effect in a semi-spherical region (dome) on top of the conductive filament, as a consequence of the spreading constraint imposed by the CMO \cite{IMWFalcone2023}. Thus, the bias-induced oxygen-related defects redistribution occurs in the same dome-shaped region where the electrothermal energy is confined, while the filament remains rigid with constant resistance (see Figure \ref{F1}a). Previous studies have outlined the material constraints on the conduction and electrothermal properties of the CMO \cite{NHFalcone2024}, which must be satisfied for the assumptions of this modeling study to hold true. Moreover, it can be extended across a range of sub-stoichiometric CMOs, including, but not limited to, the one considered in this work. \\
A compact model for analog CMO/HfO$_{\rm x}$-based ReRAM devices, accounting for the aforementioned microscopic mechanisms, has been previously introduced in \cite{ESSDERC24_Galetta} and tested in a quasi-static regime. This work demonstrates its applicability in simulating the device switching kinetics and analog pulse response, both of which have not been addressed in previous studies. 
This section reports the physical equations constituting the model core, recalling the key mechanisms occurring in the CMO during resistive switching. It also details the new modified equations to account for external parasitic resistive elements and corrected ion migration dynamics, thereby preventing unphysical concentrations of oxygen-related defects. \\

\subsubsection*{Device model}
In transition metal oxides (TMOs), like the sub-stoichiometric TaO$_{\rm x}$ acting as CMO layer in this study, the memory effect arises from the migration of oxygen anions, although their dynamics can be expressed as a motion of oxygen vacancies \cite{Waser2009}. Specifically, here only oxygen vacancies with charge equal to +2 are considered ($\rm V_{\rm O}^{\cdot \cdot}$ in Kro\"oger-Vink notation \cite{Kroger}), as they are the one with lowest formation energy \cite{Zhu2016}. Therefore, the $\rm V_{\rm O}^{\cdot \cdot}$ concentration ($C_{\rm V}$) in the CMO dome is used as a state variable to model the memory effect over the time span of the voltage input. This way, the following continuity equation describes the rate of defect displacement:

\begin{equation} 
\label{E1}
\frac{\rm d \it C_{\rm V}}{\rm d \it t} = - \frac{J_{\rm Ion}}{z q l_{\rm D}}
\end{equation}

where $z$ is the defect charge number and $q$ is the elementary charge. \\
The presence of a vertical electric field ($\mathcal{E}$) lowers the zero-field intra-lattice migration barrier ($\Delta W_{\rm A}$), establishing a preferred direction for ion drift along the dome thickness ($l_{\rm D}$). Consequently, a net motion of $\rm V_{\rm O}^{\cdot \cdot}$ arises, resulting in an ionic drift current density ($J_{\rm Ion}$), modeled as a lattice site-to-site Mott-Gurney hopping \cite{Mott_Gurney1950}. In this study, the ion drift mean velocity ($v_{\rm Ion}$) associated with the $\rm V_{\rm O}^{\cdot \cdot}$ net motion, is isolated from $J_{\rm Ion}$, enabling the analysis of the ion migration speed profile during resistive switching transitions.

\begin{align}
J_{\rm Ion} & = z q \mathcal{C}_{\rm V} \cdot v_{\rm Ion}  = \nonumber \\ & = z q \mathcal{C}_{\rm V} a \nu_{0} \cdot \exp \left(- \frac{\Delta W_{\rm A}}{k_{\rm B} T} \right) \cdot 2 \sinh \left( \frac{z q \mathcal{E} a}{2 k_{\rm B} T} \right) \cdot \left( 1 - \frac{C_{\rm V}}{C_{\rm V,lim}} \right)^{\alpha}
\label{E2-3}
\end{align}

where $a$ is the ion hopping distance, $\nu_{0}$ is the ion hopping attempt frequency, $k_{\rm B}$ is the Boltzmann constant and $T$ represents the temperature. In this work, a limiting factor to the power of $\alpha$ is introduced to ensure that $C_{\rm V}$ does not exceed unphysical values beyond $C_{\rm V,lim}$, which is determined by density of oxygen sites in the unit cell (computations reported in Supporting Information), and might be related to material phase transitions \cite{LaTorre2019}. \\
To account for the effects of parasitic resistive elements external to the ReRAM cell ($R_{\rm Series}$), as well as the voltage drops across the electrodes and the rigid, resistance-invariant conductive filament, this study introduces a voltage partitioning approach for the bias applied to the top electrode ($V_{\rm A}$). Consequently, $\mathcal{E}$ is the field portion present across the only variable resistance layer, represented by $R_{\rm CMO}$, and is, therefore, the component responsible for driving ion migration. The conductive filament and the electrodes are modeled as invariant resistors ($R_{\rm CF}$ and $2R_{\rm Ele}$ respectively), as schematically illustrated in the ReRAM equivalent circuit in Figure \ref{F1}a.

\begin{equation}
\label{E4}
R_{\rm CMO} = V_{\rm A}/I_{\rm e} - R_{\rm CF} - 2R_{\rm Ele} -R_{\rm Series}
\end{equation}

\begin{equation}
\label{E5}
\mathcal{E} = \frac{V_{\rm A}}{l_{\rm CMO}} \cdot \frac{R_{\rm CMO}}{R_{\rm CMO} +R_{\rm CF} +2R_{\rm Ele} + R_{\rm Series}}
\end{equation}

A bias-induced redistribution of ions within the dome results in variations of CMO layer conduction properties. In several TMOs, doubly charged $\rm V_{\rm O}^{\cdot \cdot}$ defects lead to the presence of donor-like levels, spectrally positioned in the middle of the band gap, and thereby, making the excitations of electrons in the conduction band energetically unfavorable \cite{Schnieders2022}. As a consequence, electron hopping through defects dominates the conduction and the average density of electronic trap states ($\mathcal{D}_{\rm e}$) dynamically modifies with alterations in $C_{\rm V}$, here simplified as:

\begin{equation}
\label{E6}
\mathcal{D}_{\rm e} = \beta z C_{\rm V}
\end{equation}

where $\beta$ is a scaling factor ($0 < \beta < 1$) representing the fraction of occupied/available states, used to filter out the donor-like levels that might not contribute to conduction. This can occur due to their misalignment with the Fermi level under applied bias, or due to electron capture by trap states. \\
Following the understanding proposed in \cite{NHFalcone2024}, the current spreading constraint makes the CMO layer the conduction bottleneck. Hence, the electronic transport flowing through the whole ReRAM device is modeled as a Trap-Assisted Tunneling (TAT) process via $\rm V_{\rm O}^{\cdot \cdot}$ deep levels in that layer, as in the following hopping form:

\begin{equation} 
\label{E7}
I_{\rm e} = A_{\rm M} q \mathcal{D}_{\rm e} a_{\rm e} \nu_{\rm e} \cdot \exp \left(- \frac{\Delta E_{\rm A}}{k_{\rm B} T} \right) \cdot 2 \sinh \left( \frac{q \mathcal{E} a_{\rm e}}{2 k_{\rm B} T} \right) 
\end{equation}

In Equation \ref{E7}, $A_{\rm M}$ represents the cross-sectional area of the CMO volume at the interface with the conductive filament, $a_{\rm e}$ is the average trap-to-trap  distance, $\nu_{\rm e}$ is the electron jump attempt frequency, and $\Delta E_{\rm A}$ is the average zero-field activation energy for TAT. \\
In addition, the lumped capacitance model is used to describe the thermodynamic phenomena confined in the CMO, as in \cite{ESSDERC24_Galetta}. The temperature transients are governed by Joule heating with respect to reference room temperature ($T_{0}$):

\begin{equation}
\label{E8}
C_{\rm th} \cdot \frac{\rm d \it T}{\rm d \it t} = P_{\rm e} - \left( \frac{T -T_{0}}{R_{\rm th}} \right)
\end{equation}

where $C_{\rm th}$ and $R_{\rm th}$ are the thermal capacitance and thermal resistance, respectively, and $P_{\rm e} = I_{\rm e} \cdot V_{\rm A}$ is the electrical power. \\
Table 1 in Supporting Information lists the simulation parameters employed to reproduce the electrothermal behavior, resistive switching characteristics, and ion dynamics of the CMO/HfO$_{\rm x}$-based ReRAM using the device model of this section.

\section{ReRAM device electrical response simulations}

\subsection{Quasi-Static $I$-$V$ Characteristic}

The previous version of the analog ReRAM compact model has been used to reproduce the quasi-static $I$-$V$ characteristic \cite{ESSDERC24_Galetta}. In this work, Equation \ref{E2-3} is employed to extend the description of the device terminal behavior to the investigation of physical properties of ion migration velocity during bipolar resistive switching, under the same quasi-static conditions as the $I$-$V$ sweep characteristic.

\begin{figure}[h]
\centering
  \includegraphics[width=8.5cm]{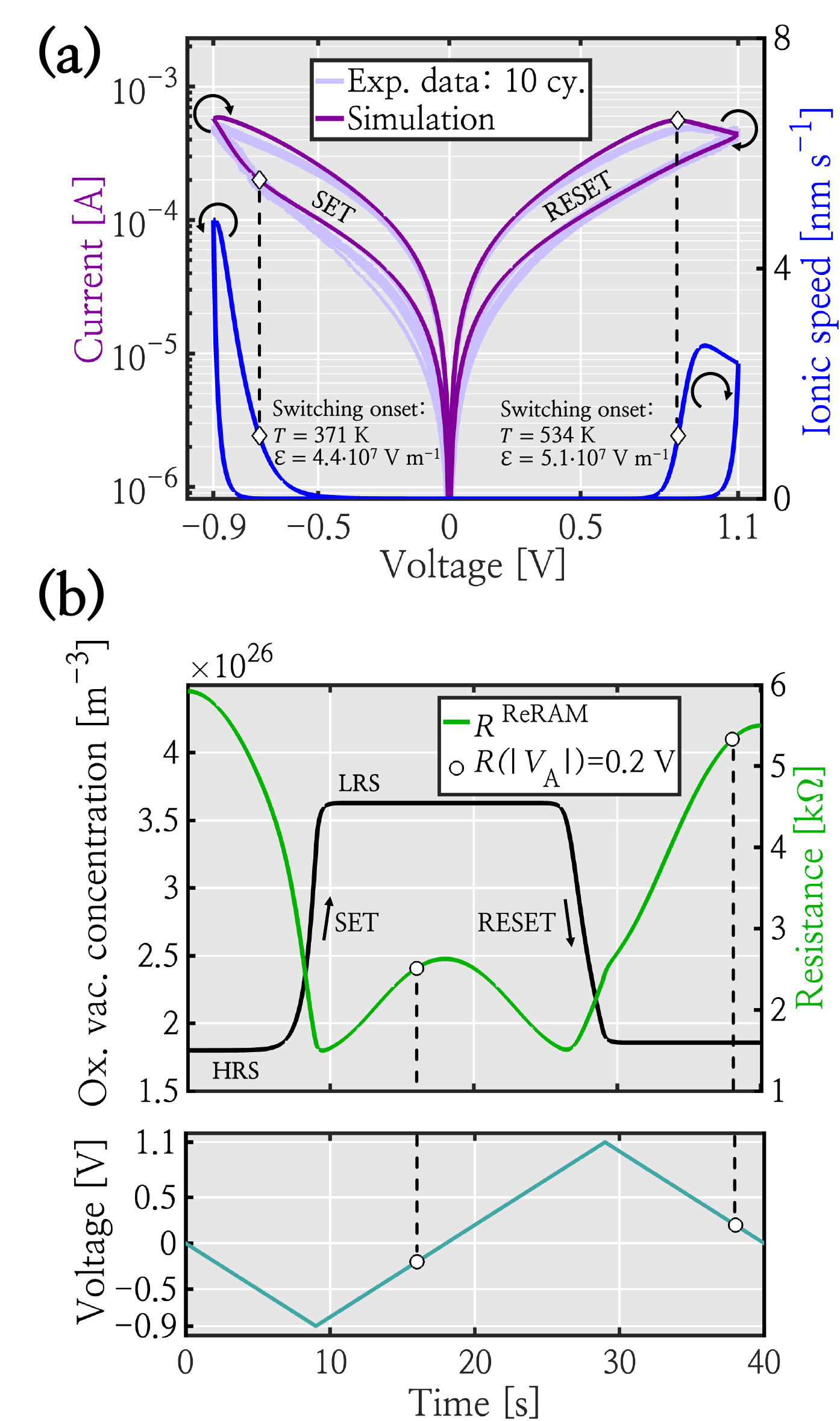}
  \caption{(a) Simulated quasi-static C8W current-voltage sweep (left axis) overlaid with experimental data (10 full sweep cycles), both obtained using the same voltage input and $R_{\rm Series} = 0 \ \Omega$ . 
  White diamonds in the current-voltage characteristic mark the SET/RESET switching onsets, which are projected along the voltage axis on the modeled ion drift velocity (right axis). (b) In the upper plot, the dynamics of oxygen vacancy concentration (left axis) and the total non-linear resistance (right axis) over the voltage input time scale. White circles intersect the resistance-time curve for $|V_{\rm A}|=0.2$ V, the voltage used to read the device state. In the bottom plot, the bipolar triangular voltage sweep ($\rm SR = 0.1$ Vs$^{-1}$) applied to the ReRAM top electrode as input to reproduce the current-voltage characteristic.}
  \label{F2}
\end{figure}

\textbf{Figure \ref{F2}}a shows 10 cycles of counter-eightwise (C8W) experimental $I$-$V$ characteristic, according to the definition reported by Dittmann et al. \cite{Dittmann2021}. The model core, including Equations \ref{E1} to \ref{E8}, is simulated with the same input as the experimental electrical conditions, i.e., the triangular voltage sweep shown in the bottom plot of Figure \ref{F2}b. The highly non-linear $I$-$V$ characteristic experimental data are accurately reproduced by a hopping transport mechanism as described with Equation \ref{E7}. Here, the intra-lattice migration barriers and the hopping conduction parameters ($a_{\rm e}, \ \Delta E_{\rm A}$) were calibrated exclusively considering the $I$-$V$ characteristic experimental data. \\
The physical framework of the compact model enables the retrieval of the ion migration velocity evolving over the applied voltage input. After validating the model's fidelity through the quasi-static $I$-$V$ characteristic, $v_{\rm Ion}$ is extracted from Equation \ref{E2-3} during the voltage sweep, and is shown in Figure \ref{F2}a. The $v_{\rm Ion}$ profile reveals an asymmetric evolution of ion migration speed between opposite switching polarities, driven by differences in electronic transport-Joule heating feedback. An exponential increase of $v_{\rm Ion}$ during the SET phase reflects the thermal runaway of the ion migration processes as a consequence of the positive feedback between conductance, temperature, and current increase. However, the partial $C_{\rm V}$ modulation leads to a gradual SET transition \cite{NHFalcone2024}, in contrast to the abrupt SET observed in metal-oxide ReRAM exhibiting a filament gap formation \cite{Larentis2012}. Conversely, during the RESET phase, the ion migration is slowed down (see Figure \ref{F2}a), as Joule heating is counteracted by a current and conductance decrease. Despite the intrinsic electrothermal asymmetry of bipolar switching dynamics (see $\mathcal{E}$ and $T$ at the switching onset in Figure \ref{F2}a) \cite{Dittmann2021}, the ionic mean speed coincides for both SET and RESET transitions at the switching onset. The value of $v_{\rm Ion}$ initiating the migration ($\sim 1.1$ nm s$^{-1}$) is found by projecting the switching onset of the simulated $I$-$V$ characteristic onto the ionic mean speed profile. Consequently, the average migration displacement is computed over a time interval ($\Delta t_{\rm Sw}$) from the switching onset to the stop voltage in the triangular sweep as:

\begin{equation} 
\label{E9}
<s_{\rm M}> \ = \int_{\Delta t_{\rm Sw}} v_{\rm Ion} \, \rm d \it t
\end{equation}

For both SET and RESET transitions, $<s_{\rm M}> \ \sim 6$ nm, regardless of electrothermal disparities. This suggests that the resistive switching in CMO/HfO$_{\rm x}$-based ReRAM is governed by bidirectional migration of the same ionic species within a confined region thinner than $l_{\rm D}$.
However, more rigorous and advanced spectroscopy techniques are required to experimentally confirm variations in defect concentration within the dome portion. 
As a result of $\rm V_{\rm O}^{\cdot \cdot}$ migration, the CMO dome undergoes a modulation of defect concentration, affecting the density of electron trap states, as described in Equation \ref{E6}. Figure \ref{F2}b shows that the CMO is partially depleted of $\rm V_{\rm O}^{\cdot \cdot}$ in HRS and populated back in LRS ($C_{\rm V}^{\rm HRS} < C_{\rm V}^{\rm LRS}$). The assumed 3D spatial distribution of doubly-charged oxygen vacancies within the CMO/HfO$_{\rm x}$ layer stack for different resistive states is schematically represented in \textbf{Figure S1} in Supporting Information. In LRS, the homogeneously distributed traps within the dome allow electron jumps among regular TAT paths (minimum $a_{\rm e}, \ \Delta E_{\rm A}$). Whereas during RESET, $\rm V_{\rm O}^{\cdot \cdot}$ migrate towards the interface with the conductive filament \cite{NHFalcone2024}, limiting the conduction in HRS by a trap deficit in the dome (maximum $a_{\rm e}, \ \Delta E_{\rm A}$) \cite{Graves2017}.
To model the gradual conductivity change, a parametrization of the average trap-to-trap distance and the TAT activation energy is considered. The parameters $a_{\rm e}, \ \Delta E_{\rm A}$ are modeled as linearly increasing with the decrease of the defect concentration in the dome \cite{Goldfarb2012}:

\begin{align}
\label{E10a}
a_{\rm e} & = a_{\rm e,0} -k^{'} \cdot C_{\rm V}  \\
\Delta E_{\rm A} & = \Delta E_{\rm A,0} -k^{''} \cdot C_{\rm V}
\label{E10b}
\end{align}

The ReRAM device experiences SET and RESET transitions when the electrothermal energy is sufficiently high to overcome the ion migration barrier. Here, $\Delta W_{\rm A}$ grows with $C_{\rm V}$ throughout the SET transition \cite{ESSDERC24_Galetta}, as the migration of $\rm V_{\rm O}^{\cdot \cdot}$ is thermodynamically favored by the strong gradient of defect concentration at the CMO/conductive filament interface \cite{Woo2016}. \\
Although during the non-switching phase $C_{\rm V}$ remains constant within the dome, the resistance of the device varies dynamically with the applied voltage, as shown in Figure \ref{F2}b. This behavior arises from the inherently non-linear TAT transport: Joule heating and the applied bias accelerate electron jumps among trap states, leading to variations in resistance without any $\Delta C_{\rm V}$ induced by ion migration. Hereby, memory states (i.e., device conductance or resistance levels) are retrieved as a combination of reading voltage and $C_{\rm V}$, with the latter uniquely identifying the device configuration (i.e., defect state). $|V_{\rm A}|=0.2$ V is chosen as a low bias value to read the memory state, when electrons are not strongly accelerated and the $I$-$V$ relationship is nearly linear.

\subsection{Switching Kinetics and Temperature-Programming Time Trade-Off}

Switching kinetics in ReRAM devices is known to be highly non-linear, especially when resistive switching is driven by thermally accelerated ionic drift \cite{Menzel2011}. As a consequence, the time of programming experiences a non-linear voltage dependency.
A strong non-linearity in switching kinetics is beneficial for achieving fast, non-volatile switching and long-term retention, as it prevents perturbation of the device state upon repeated low-voltage read pulses that are too weak to trigger switching transitions. In addition, this pronounced non-linearity implies programming of memory states using short voltage pulses, thereby addressing what is commonly referred to in the literature as the voltage-time dilemma \cite{Waser2009}. \\

\begin{figure}[H]
\centering
  \includegraphics[width=18cm]{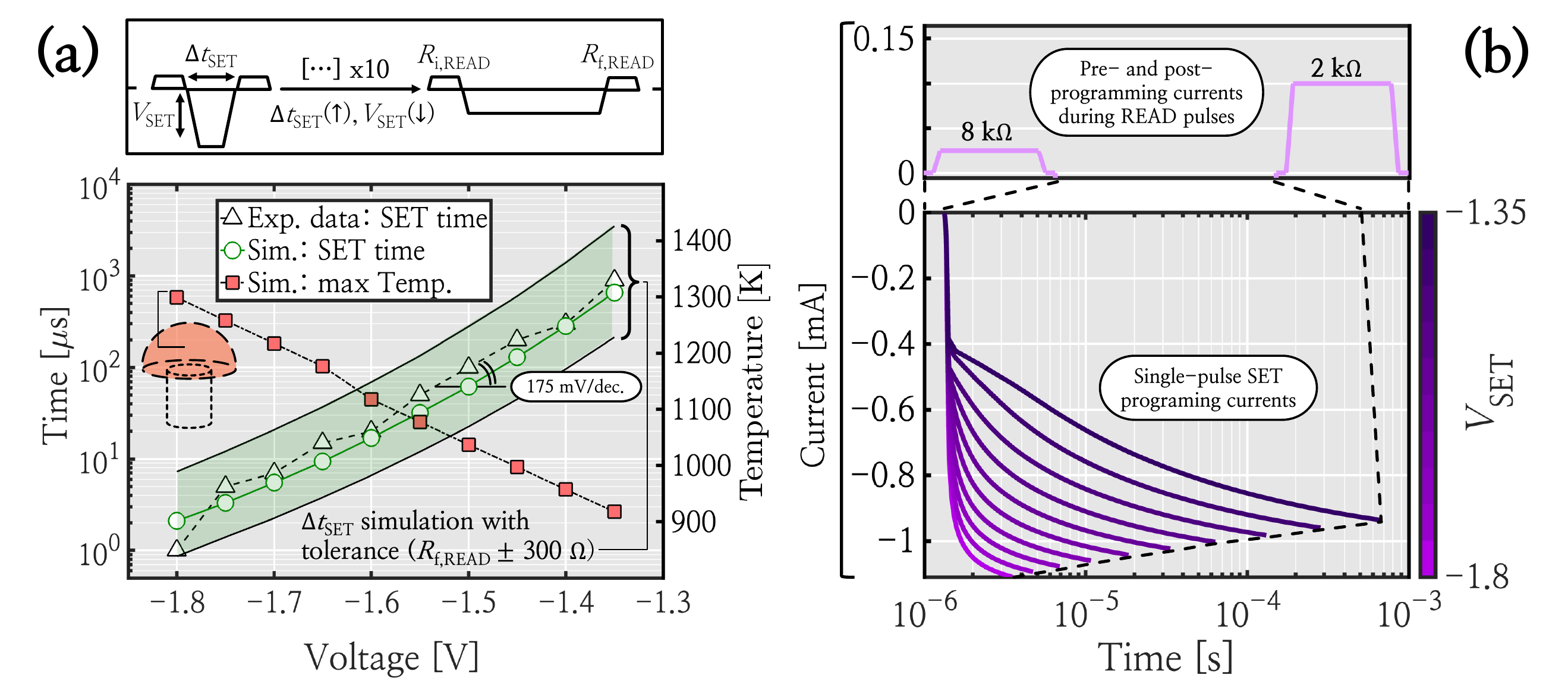}
  \caption{(a) In the top panel, a sketch of the READ-SET-READ pulse sequence used to study the switching kinetics through single-pulse programming. In the bottom plot, the time-voltage ($\Delta t_{\rm SET}$-$V_{\rm SET}$) trade-off for single-pulse programming, overlaying experimental and simulated SET times (left axis), along with the maximum temperature reached within the switching volume (right axis) as a function of pulse amplitude. The maximum temperature refers to the peak value during the SET programming pulse. Experimental and simulated voltages are applied to a series including the ReRAM and a load resistor ($R_{\rm Series} = 50 \ \Omega$) used for current measurement in the experimental setup. The simulation is carried out with an initial oxygen vacancies concentration of $C_{\rm V,0} = 1.37 \cdot 10^{26}$ m$^{-3}$, corresponding to $R_{\rm i,READ} =$ 8 k$\Omega$, when reading at $|V_{\rm A}|=0.2$ V. (b) Current transients over time within the programming pulse width $\Delta t_{\rm SET}$ for $V_{\rm SET}$ values ranging from -1.35 V to -1.8 V. Each current transient corresponds to a simulated SET pulse (white circles) that satisfies the criterion for pre- and post-programming resistance ($R_{\rm i,READ} =$ 8 k$\Omega \rightarrow R_{\rm f,READ} =$ 2 k$\Omega$).}
  \label{F3}
\end{figure}

The SET switching kinetics of the CMO/HfO$_{\rm x}$-based ReRAM is characterized and simulated in this section, with the read voltage ($V_{\rm READ}$) fixed at $|V_{\rm A}|=0.2$ V to ensure consistency throughout this work. SET programming voltages that do not exceed approximately $10 \cdot V_{\rm READ}$ \cite{Dittmann2021} are used throughout the experiment.
\textbf{Figure \ref{F3}}a shows the exponential dependency of the SET switching time ($\Delta t_{\rm SET}$) on $V_{\rm SET}$ within the selected programming voltage range. To characterize the SET switching kinetics, single-pulse programming data are collected under fixed switching conditions, with the pre- and post-programming resistance values set to $R_{\rm i,READ}= 8$ k$\Omega$ and $R_{\rm f,READ}= 2$ k$\Omega$, respectively. A tolerance of $\pm 300 \ \Omega$ is allowed for reaching $R_{\rm f,READ}$. These values define the programming criterion used to consider an experimental data point as valid.
Equally with the experimental programming conditions, identical READ-SET-READ pulse sequences are reproduced as input for the simulation. Figure \ref{F3}a shows agreement between the experimental and simulated switching kinetics, confirming that the model core accurately predicts resistive switching behavior across a wide range of voltage and time programming conditions. Besides characterizing the switching kinetics, Figure \ref{F3}a enables the design of the appropriate pulse amplitude and duration to program the memory cell in a binary fashion by a single square voltage pulse, without accessing the IRSs.

An improper balance of $V_{\rm SET}$ and $\Delta t_{\rm SET}$ can lead to incomplete switching or excessive Joule heating/device degradation (region below or above the $\Delta t_{\rm SET}(V_{\rm SET})$ curve respectively). A single slope in the $\Delta t_{\rm SET}(V_{\rm SET})$ diagram, plotted on a logarithmic scale, denotes Joule heating as the dominant driving mechanism accelerating the memory state switching \cite{Menzel2015}. This behavior is typically associated with ion migration strongly confined within a region of few nanometers, supporting the average migration displacement estimated by Equation \ref{E9}. Consequently, the high non-linearity of switching kinetics ($\sim$$175$ mV/dec) in CMO/HfO$_{\rm x}$-based ReRAM might be attributed to the electrothermal confinement in the nanometer-scale dome region on top of the conductive filament, where the ion mobility is enhanced by Joule heating. Figure \ref{F3}b shows the simulated electronic current during the SET transition for each pulse, used to compute the electrical energy of programming ($E_{\rm p,SET}$) as:

\begin{equation} 
\label{E12}
E_{\rm p,SET} = \int_{ \Delta t_{\rm SET}} V_{\rm SET} \cdot I_{\rm e} \, \rm d \it t
\end{equation}

The strong non-linearity of the $\Delta t_{\rm SET}(V_{\rm SET})$ diagram reflects in the voltage-dependency of $E_{\rm p,SET}$: large programming voltages with short pulses enable low energy switching (see \textbf{Figure S3} in Supporting Information) at the expense of a large maximum temperature within the switching region in the CMO. Despite fast and low energy switching is preferred to outperform other memory technologies (DRAM or high performance SRAM \cite{Waser2009}), a temperature-programming time trade-off has to be considered to avoid performance degradation of CMO/HfO$_{\rm x}$-based ReRAM attributed to excessive Joule heating.

\subsection{Analog Bidirectional Accumulative Conductance Response}\label{ABAC}

\begin{figure}[H]
\centering
  \includegraphics[width=8cm]{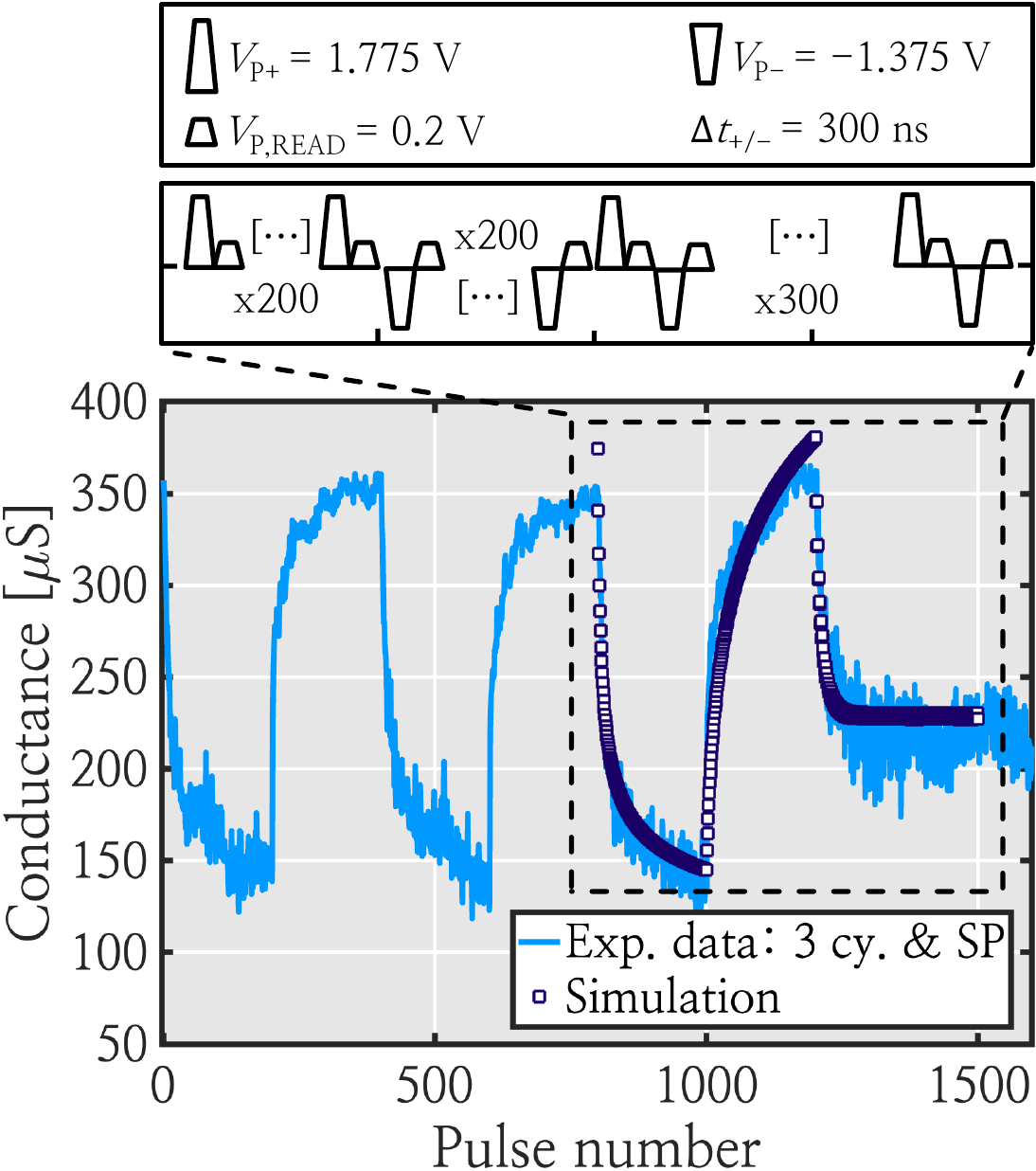}
  \caption{Analog bidirectional accumulative conductance response of the ReRAM undergoing bipolar batches of voltage pulses. In the top panel, the experimental and simulated pulse scheme for 1 conductance depression/potentiation cycle ($V_{\rm P+}$/$V_{\rm P-}$ as pulse amplitudes and $\Delta t_{+/-}$ as pulse duration) and a sequence of alternating $V_{\rm P+}$/$V_{\rm P-}$ pulses to catch the symmetry point memory state. The simulation is carried out with an initial oxygen vacancies concentration of $C_{\rm V,0} = 3.15 \cdot 10^{26}$ m$^{-3}$ for the first simulated data point and $R_{\rm Series} = 50 \ \Omega$, as in the experiment for current sensing.}
  \label{F4}
\end{figure}

On-chip training using analog cross-point architectures was shown to be feasible by implementing hardware-aware algorithms, namely Tiki-Taka \cite{Gokmen2020}. Such algorithms account for device non-linear switching response and assume that cross-point memory cells exhibit a SP and analog bidirectional accumulative conductance updates.  
Thereby, the cross-point ReRAM devices are studied to meet such criteria when tested under bipolar voltage pulse streams, where each batch involves pulses identical in amplitude and duration. Specifically, CMO/HfO$_{\rm x}$-based ReRAM devices exhibit analog bidirectional resistive switching, where each pulse leads to a small conductance ($G$) change, therefore allowing for access to multiple analog memory states. Furthermore, upon programming with alternating positive/negative pulses, the device features a given SP within the $G$-window \cite{Stecconi2024}. In this study, the analog bidirectional accumulative conductance response of CMO/HfO$_{\rm x}$-based ReRAM is evaluated experimentally and simulated via physics-aware compact model, under Tiki-Taka training pulsing scenario conditions. To this aim, a predefined pulse scheme is applied, without involving a real-time feedback after each pulse to adjust the programming procedure for catching the desired device memory state. Such approach, also known as open-loop programming, consists of applying a batch of 200 positive ($V_{\rm P+}$) and 200 negative pulses ($V_{\rm P-}$) with duration of $\Delta t_{+/-} = 300$ ns to gradually decrease (down) and increase (up) the conductance respectively within a stable $G$-window.
300 alternating positive/negative pulses are used to program the ReRAM around the SP. Read pulses of $V_{\rm P,READ}=0.2$ V and long 200 ns, are used to evaluate the device conductance after each programming pulse of the sequence. 
Experimental data reported in \textbf{Figure \ref{F4}} show the analog bidirectional accumulative conductance response over 3 cycles of down/up and 1 SP procedure, selected based on reproducible behavior within a stable $G$-window observed across multiple devices (see \textbf{Figure S4} in Supporting Information).

Open-loop programming of the device conductance and the oscillating behavior of the physical system around the SP are closely reproduced by the simulation model. Throughout the up/down phases, each programming pulse triggers ion displacement depending on the voltage pulse polarity. This results in a cumulative change in $C_{\rm V}$. Therefore, IRSs across the $G$-window can be retrieved through variations in the available electron trap states ($\mathcal{D}_{\rm e} \propto C_{\rm V}$) during the programming pulse duration, while remaining constant upon reading:

\begin{equation} 
\label{E13}
G_{\rm READ} = \frac{I_{\rm e}}{V_{\rm P,READ}} \propto \mathcal{D}_{\rm e}
\end{equation}

\begin{figure}[H]
\centering
  \includegraphics[width=18cm]{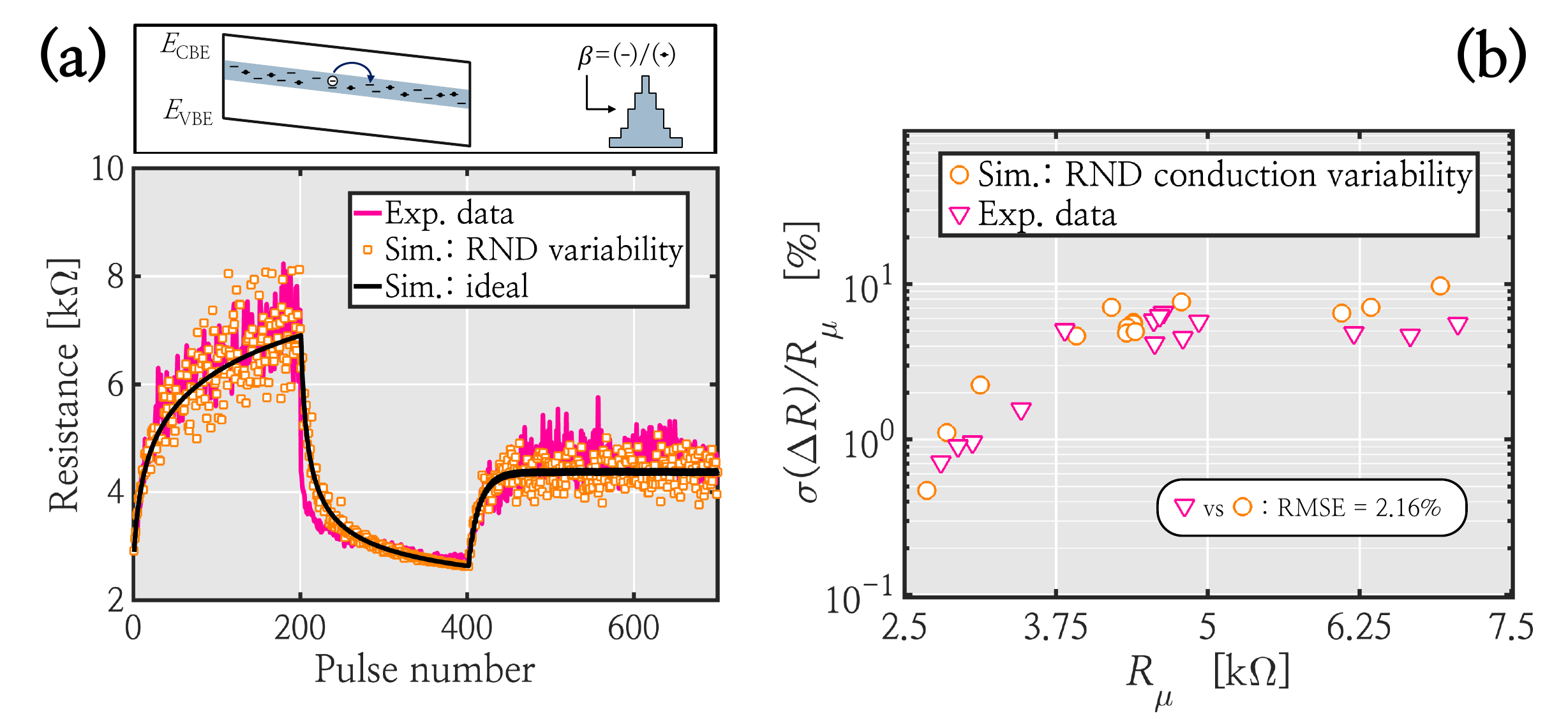}
  \caption{(a) Analog accumulative response, simulated including conduction variability as fluctuations in the ratio between free/occupied trap states for electrons hopping sampled from a random distribution. The top panel includes a sketch of the physical mechanisms responsible for conduction fluctuations. Free/occupied traps are schematically represented as non-dotted/dotted lines spectrally located in the middle of the band gap, and their fraction randomly changes as Gaussian dispersion. The simulation is carried out under the same conditions as in Figure \ref{F4}. (b) Normalized Root-Mean-Square Deviation comparing the dispersion of resistance fluctuations among experimental and simulated cumulative response characteristics.}
  \label{F5}
\end{figure}

Parameter variability is incorporated into the simulation model to account for the stochasticity of the conductance levels during open-loop programming. In similar compact models for valence-change memories, physical noise on the read resistance is tied to the number of $\rm V_{\rm O}^{\cdot \cdot}$ within the switching layer \cite{Wiefels2020, Bengel2020}. In the presented model, $G_{\rm READ}$ is directly proportional to $\mathcal{D}_{\rm e}$. Hence, its value can be modeled as the source of conduction variability. Under the circumstances of defects position change, abrupt current fluctuations are typically observed \cite{Schnieders2022}. Ion migration can cause irregularities in the hopping paths through defect trap states, influencing the population of electrons. Accordingly, the fraction of occupied/available electron trap states—represented by the scaling factor $\beta$—might change irregularly between two different programmed states, resulting in current fluctuations. Moreover, when fewer defects govern conduction, current fluctuations become more pronounced, leading to greater dispersion \cite{Dittmann2021}. Random fluctuations of the physical parameter $\beta$ are simulated by sampling its value from a Gaussian distribution having the default $\beta$-parameter of the variability-free simulation as its mean value ($\mu_{\beta}$). The standard deviation ($\sigma_{\beta}$) is modeled as linearly inversely dependent on the defect state to account for the increased fluctuations observed at higher resistance values:

\begin{equation} 
\label{E13b}
\sigma_{\beta} = \sigma_{\beta,0} -k^{'''} \cdot C_{\rm V}
\end{equation}

where $\sigma_{\beta,0}$, $k^{'''}$ have been treated as model parameter, and calibrated to fit the experimental data.  \textbf{Figure \ref{F5}}a shows the simulated bidirectional accumulative resistance response, including the random model for the $\beta$-parameter to fit the experimental resistance dispersion. This approach simplifies the trap-assisted electron capture and emission process by modeling it as a stochastic TAT mechanism without explicit time constants. 
The Normalized Root-Mean-Square Deviation (NRMSD) is computed to assess the affinity of experimental and simulated resistance fluctuation dispersion. Figure \ref{F5}b shows the NRMSD as a function of the mean resistance ($R_{\mu}$), highlighting a strong agreement between the stochastic model and the measured trend (2.16\% Root Mean Squared Error between the two trends). In Figure \ref{F5}b, each circle refers to a batch of 50 consecutive read resistance points out of the total 700 pulses.

As a result, this section demonstrates the capability of the advanced analog ReRAM compact model to accurately reproduce realistic analog resistive switching characteristics, including conduction variability.

\section{Analysis of Memory States Equilibria}

\subsection{Physical Origin of the symmetry point}
Recent studies on resistive switching in ReRAMs, triggered by periodic voltage stimuli, have reported an oscillatory response around memory states due to SET/RESET equilibrium \cite{Ascoli2022}. Additionally, when a physical system exhibits a unique asymptotic behavior that is independent of initial conditions for a given electrical input, it is referred to as a fading memory mechanism. 
This study investigates the equilibrium of memory and defect states in CMO/HfO$_{\rm x}$-based ReRAM and their evolution over time, to establish a correlation between the physical criterion satisfied at the equilibrium point under quasi-static switching conditions and under pulsed response around the SP. To this end, the Dynamic Route Map (DRM) approach presented in \cite{Ascoli2022, Ascoli2024} is tested with the simulation model provided in section \ref{CM}. From a purely mathematical perspective, the DRM for a generic simulation model of a physical system consists of a class of State Dynamic Routes (SDRs). Each SDR represents the relationship between the time derivative of the state variable (or state variable flux, $\Phi$) and the state variable itself. In the case of the CMO/HfO$_{\rm x}$-based ReRAM model, the device configuration is identified by the defect state $C_{\rm V}$. Hence, SDRs describing the SET or RESET dynamics are derived from Equation \ref{E1} as a function of $C_{\rm V}$.

\begin{equation}
\label{E14}
\Phi = \frac{\rm d \it C_{\rm V}}{\rm d \it t} = - \frac{J_{\rm Ion}}{z q l_{\rm D}}
\end{equation}

\begin{figure}[H]
\centering
  \includegraphics[width=15cm]{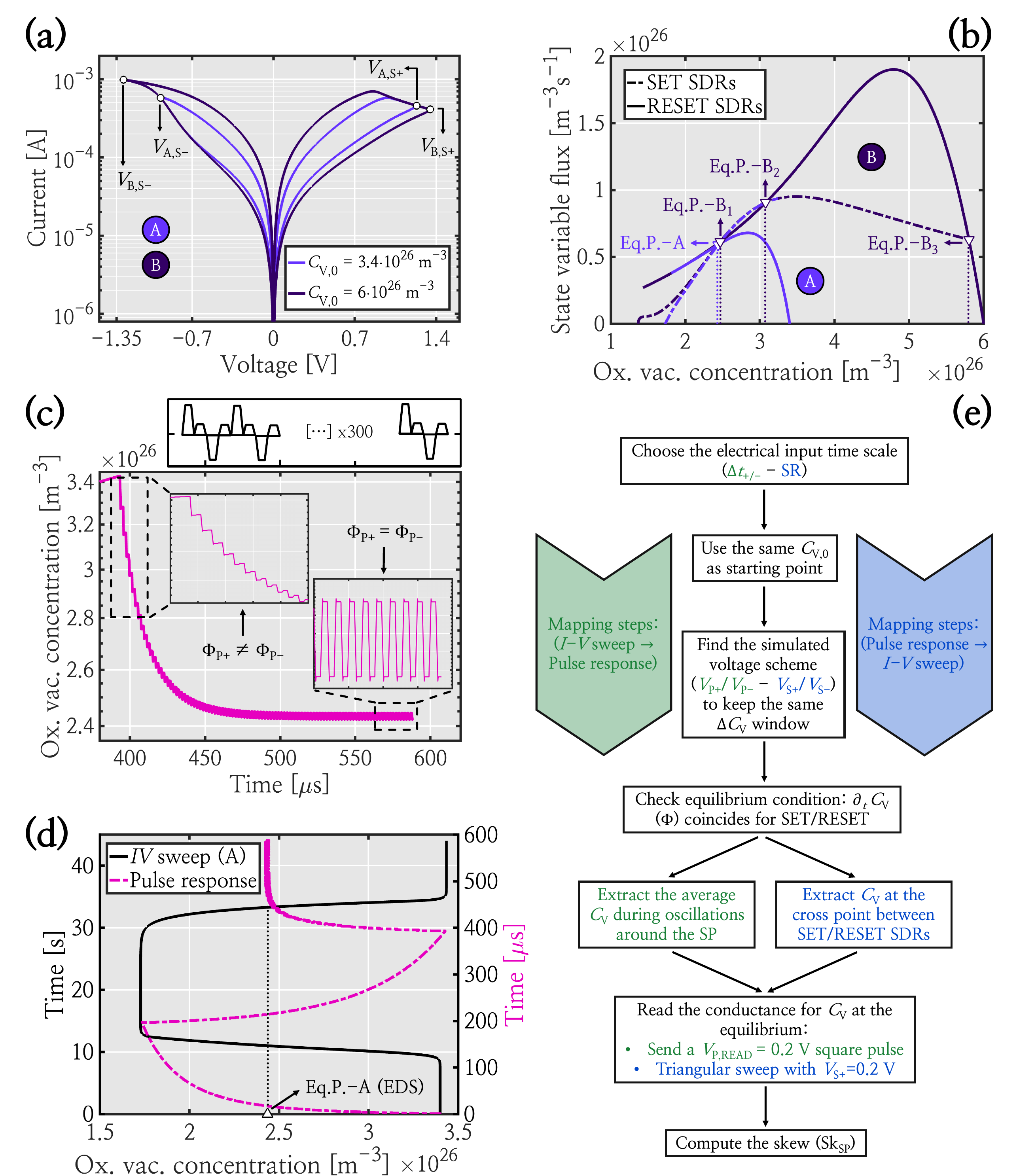}
  \caption{(a) Quasi-static current-voltage sweeps (cases A \& B) simulated with varying positive/negative stop voltages ($V_{\rm S+}$/$V_{\rm S-}$) and conductance window to investigate the equilibria of memory states. (b) State Dynamic Routes are divided for SET/RESET (dashed/solid lines) phases in each A \& B case. Cross-points between SET/RESET dynamics reveal the equilibrium points of device memory states (marked with triangles). (c) Temporal dynamics of oxygen vacancy concentration when the device undergoes the periodic voltage input sketched in the top panel. The left inset highlights the unbalanced change of oxygen vacancy concentration. While fading memory mechanism occurrence is shown in the right inset, around a constant stable memory state. (d) Overlay of the temporal dynamics of oxygen vacancy concentration. The device memory state is tuned with quasi-static voltage sweep input (left axis) and with bipolar batches of voltage pulses (right axis). The black dotted line shows the coincidence between the equilibrium defect state found through quasi-static Static Dynamic Routes and the symmetry point defect state of analog accumulative response within the same $\Delta C_{\rm V}$ window. (e) Schematic procedure to map the equilibrium defect state into the symmetry point defect state across input domains, i.e., quasi-static voltage sweep and accumulative pulse response. Green steps allows to retrieve a bidirectional accumulative pulse response having the symmetry point that coincides with the equilibrium defect state read at $|V_{\rm A}|=0.2$ V. Blue steps allow to retrieve a quasi-static $I$-$V$ sweep having the equilibrium defect state read at $|V_{\rm A}|=0.2$ V that coincides with the symmetry point.}
  \label{F6}
\end{figure}

The study under quasi-static conditions is conducted on two distinct cases (A \& B), distinguished by the stop voltages of the input triangular sweeps ($V_{\rm B,S+} > V_{\rm A,S+} \ \& \ V_{\rm B,S-} < V_{\rm A,S-}$) and by the initial oxygen vacancies concentration ($C_{\rm V,0}$), i.e. the initial defect state. \textbf{Figure \ref{F6}}a depicts the simulated quasi-static $I$-$V$ sweeps under A,B conditions. Whereas Figure \ref{F6}b shows the DRM related to the switching phases only, i.e., when there is a significant change in the defect state. In case A, the SET and RESET SDRs intersect at a single point, identifying the Equilibrium Point (Eq.P.-A) where the state variable flux is equal for both switching transitions. 
Although ion migration processes during SET and RESET are electrothermally asymmetric due to the nature of the electrothermal feedback, at Eq.P.-A, the driving force for $\rm V_{\rm O}^{\cdot \cdot}$ migration is equal in both directions. Therefore, $C_{\rm V}$ at Eq.P.-A is referred to as the Equilibrium Defect State (EDS).
It is worth noting that the ReRAM device does not exhibit one unique EDS: by expanding the resistive window, multiple equilibrium points can emerge, as shown in Figure \ref{F6}b with the SDRs for case B (Eq.P.-B$_{1,2,3}$). Furthermore, the uniqueness of each EDS is violated if the electrical input changes, as demonstrated by the inequality Eq.P.-A $\neq$ Eq.P.-B$_{1}$, as different triangular sweeps are used when simulating case A and case B. \\

During analog bidirectional accumulative conductance response evaluation, a periodic electrical input of alternating positive/negative pulses is used to program the ReRAM around the SP. The temporal evolution of the defect state throughout the SP phase (300 alternating $V_{\rm P+}, \ V_{\rm P-}$ pulses) is shown in Figure \ref{F6}c, simulated under the same initial and pulsing conditions as in Figure \ref{F4}. Following a transient phase while alternating $V_{\rm P+}$ and $V_{\rm P-}$ imply non-equal changes in $C_{\rm V}$, the strengths of pulses with opposite polarities converge and become equal, i.e. the ReRAM is programmed around the SP. The SP is reached independently of the initial conditions for a given input, as demonstrated by \textbf{Figure S5} in Supporting Information. Thereby, during the SP phase, the ReRAM exhibits a fading memory mechanism, erasing the memory of the starting level. Equation \ref{E14} is used to compute the state variable flux for each programming pulse in the course of the SP phase. $V_{\rm P+}$ and $V_{\rm P-}$ result in different or equal $\Phi_{\rm P+}$, $\Phi_{\rm P-}$ in the transient phase or once the SP is reached, respectively (see Figure \ref{F6}c insets). It can be concluded that the origin of the SP during the occurrence of the fading memory mechanism is tied to the same equilibrium criterion satisfied around the EDS under quasi-static operation. In both cases, electrothermally asymmetric bidirectional ion motions are rebalanced by external electrical stimuli.

To assess the equivalence of the equilibrium condition, the quasi-static and pulse responses are studied together. The $C_{\rm V}(t)$ curves related to the two inputs are plotted in Figure \ref{F6}d: the pulse response refers to the same input of Figure \ref{F4}, while the quasi-static one corresponds to the case A of Figure \ref{F6}a. Both scenarios result in an identical memory state window when starting from the same stage. Figure \ref{F6}d reveals that the EDS found through the DRM approach for quasi-static conditions matches the defect state of the SP. Therefore, given the same memory state window, there exists at least one combination of external stimuli that results in equilibrium conditions around the same defect state, despite variations in the electrical input. These findings reveal the conditions under which the same equilibrium point manifests across different input domains (i.e., quasi-static voltage sweep and accumulative pulse responses), thus enabling the identification of the SP from an $I$-$V$ sweep—or an $I$-$V$ sweep featuring a stable resistance window for a given SP.

\subsection{Memory State Mapping Across Input Domains}

In the context of device criteria for the implementation of the Tiki-Taka algorithm, the switching symmetry within a $G$-window is evaluated through the SP skew ($\rm Sk_{\rm SP}$) \cite{Stecconi2024}: it is quantified in percentage, with 50\% corresponding to a perfect symmetry between potentiation and depression phases, therefore having the SP in the middle of the $G$-window. Here, a skew mapping procedure is built upon the equivalence between the EDS, identified by the DRM approach under quasi-static conditions, and the SP defect state. Figure \ref{F6}e illustrates the reversible mapping across input domains, summarizing the steps followed over the simulations related to Figure \ref{F6}d. Additionally, once the $C_{\rm V}$ corresponding either to the EDS or the SP defect state is found, $\rm Sk_{\rm SP}$ can be computed from both quasi-static $I$-$V$ sweep or pulse response simulations according to:

\begin{equation}
\label{E15}
\rm Sk_{\rm SP} = \frac{\it G_{\rm LRS} - \it G_{\rm EDS}}{\it  G_{\rm LRS} -\it G_{\rm HRS}} \leftrightarrow \rm Sk_{\rm SP} = \frac{\it G_{\rm max} -\it G_{\rm SP}}{\it G_{\rm max} -\it G_{\rm min}}
\end{equation}

\begin{figure}[H]
\centering
  \includegraphics[width=15cm]{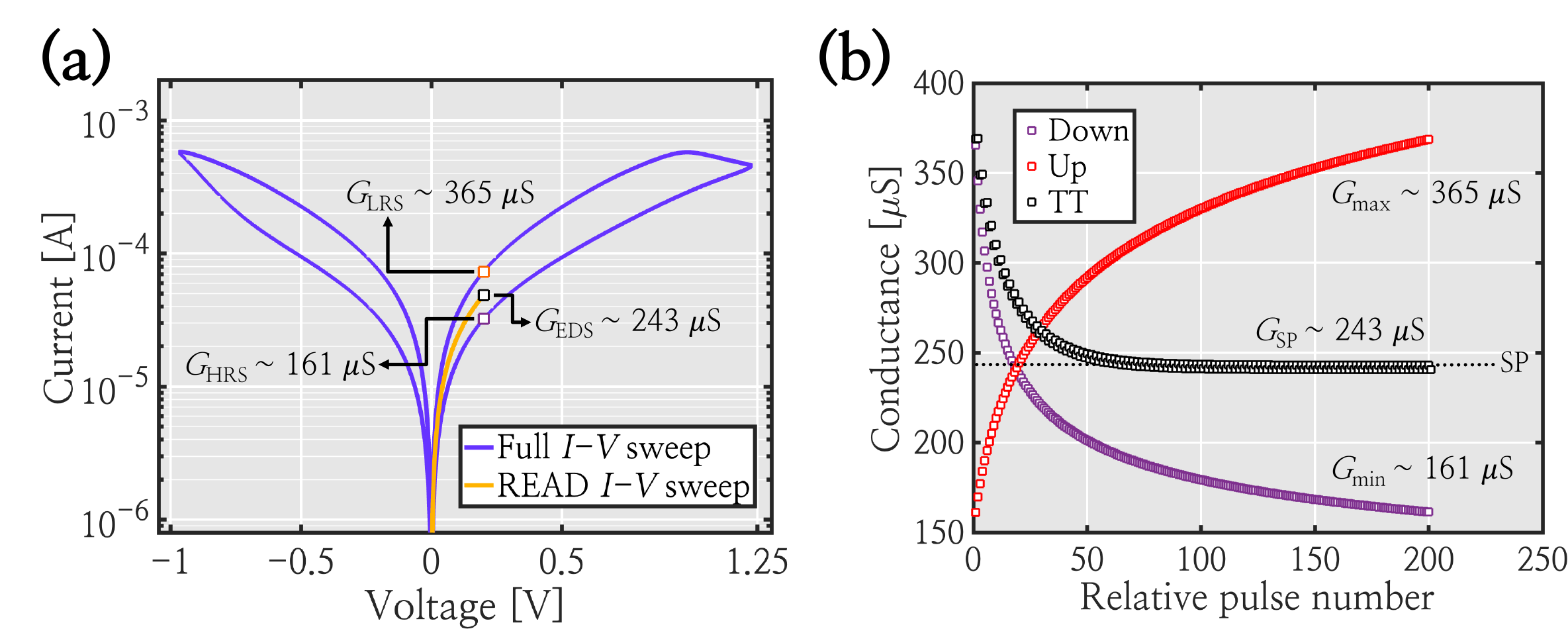}
  \caption{(a) Quasi-static current-voltage sweep (case A) taken as a reference to test the skew mapping procedure. $G_{\rm HRS}$/$G_{\rm LRS}$/$G_{\rm EDS}$ are the parameters used to compute the correctly mapped $\rm Sk_{\rm SP}$ in quasi-static regime. (b) Bidirectional accumulative analog conductance response in the same conductance window of quasi-static current-voltage sweep (case A). $\Delta t_{+/-}=300$ ns, $V_{\rm P+}=1.706$ V, $V_{\rm P-}=-1.38$ V are used, following the green steps of the skew mapping procedure. $G_{\rm min}$/$G_{\rm max}$/$G_{\rm SP}$ are used to compute Sk$_{\rm SP}$.}
  \label{F7}
\end{figure}

\textbf{Figure \ref{F7}}a and Figure \ref{F7}b show the quasi-static $I$-$V$ sweep and the accumulative conductance response used to test the skew mapping procedure. Figure \ref{F7}a and Figure \ref{F7}b are the results of blue and green steps illustrated in Figure \ref{F6}e, respectively.

\section{Training Simulations with Tiki-Taka Algorithm through Pulse-Amplitude Tuning}

\begin{figure}[h]
\centering
  \includegraphics[width=16.5cm]{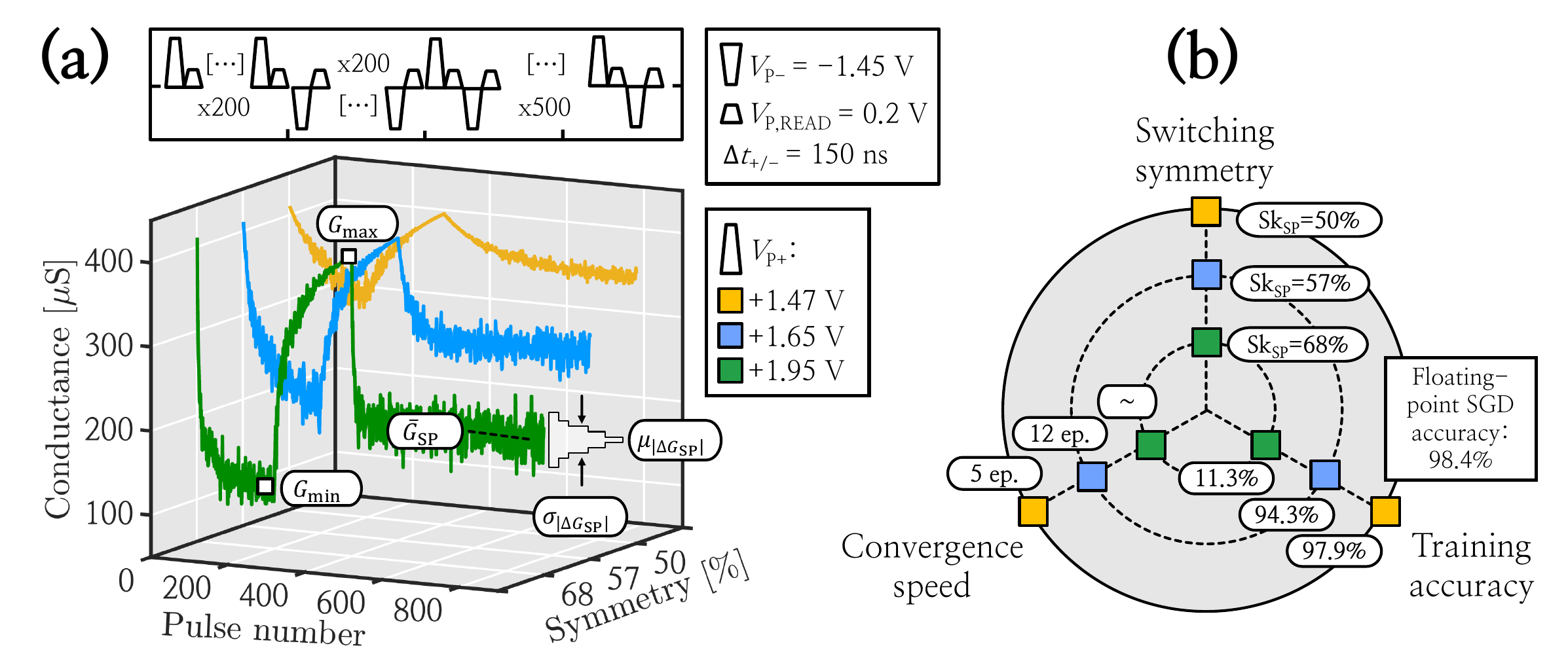}
  \caption{(a) Simulated traces of analog bidirectional accumulative conductance response used to evaluate the impact of the switching characteristic symmetry on training performance. (b) Training simulation results on a 3-FC (784-256-128-10) network for MNIST digit classification.  For the floating-point SGD reference, the weight matrix was updated with a learning rate 'lr = 0.01'. For TTv1, training simulations were done with ‘fast\_lr’ = 0.1 for gradient accumulation in matrix (A) and transferring to weight matrix (C) every ‘transfer\_every’ = 3 times. Effect of tuning the $V_{\rm P+}$ on Tiki-Taka training accuracy is based on two key performance metrics, i.e., convergence speed (in number of epochs), and final training accuracy with respect to the switching symmetry (Sk$_{\rm SP}$).
}
\label{F8}
\end{figure}

Analog training based on the Tiki-Taka algorithm (TTv1) relaxes device constraints typically required for Stochastic Gradient Descent (SGD) convergence \cite{Gokmen2020, Rasch2020Training} by introducing an additional matrix, A. This matrix, used for gradient accumulation, involves ReRAM devices operated around the SP. However, to reach high training accuracy, some fundamental device requirements remain in terms of the number of stable conductance states, a Noise-to-Signal Ratio $< 100\%$ and most importantly, a centered SP \cite{Stecconi2024}. In an updated version of Tiki-Taka (TTv2), the requirements of the number of reliable conductance states were considerably relaxed \cite{Gokmen2021} by introducing a digital low-pass filtering stage, although still relying on the accurate programming of the SP in matrix A. The latest and more refined version, i.e., Analog Gradient Accumulation with Dynamic reference (AGAD or TTv4), strengthens the resilience of the algorithm to non-symmetrical switching characteristics (i.e,. non-centered SP) by estimating the SP iteratively \cite{Rasch2024Agad}. This is achieved through additional off-chip, on-the-fly digital computations, which incur an added computational complexity cost \cite{Rasch2024Agad}. \\
In this section, we investigate how tuning the pulse amplitudes—within the physics-based compact model simulation—affects the symmetry of the switching characteristic, impacting the analog training performance when employing the earliest version of Tiki-Taka, TTv1. \\
The analog CMO/HfO$_{\rm x}$-based ReRAM compact model used in this study has been tested for devices exhibiting a slightly more gradual potentiation compared to the depression phase, therefore having a SP below the center of the conductance window (Sk$_{\rm SP} > 50\%$). However, the SP is determined by the strength of the device's response upon alternating pulses ($V_{\rm P+}$ and $V_{\rm P-}$) and its memory state can be tuned accordingly \cite{abedin2023}. Therefore, the pulse amplitude during the depression phase is freely adjusted, resulting in different percentages of symmetry (Sk$_{\rm SP} = \{50\%, \ 57\%, \ 68\% \}$ ). \textbf{Figure \ref{F8}}a shows three generated traces of analog bidirectional accumulative conductance response, varying the strength of $V_{\rm P+}$ (+1.47 V, +1.65 V, and +1.95 V) in the compact model simulation. \\
The individual traces are averaged and max-normalized to the (-1, 1) range, and posteriorly fitted to a \textit{soft-bound} analytical device model \cite{Frascaroli2018} in the ’aihwkit’ \cite{aihwkit}. The \textit{soft-bound} device realistically reproduces the switching characteristics for weight updates during analog training. The three \textit{soft-bound} device presets account for the conductance range, minimum conductance step, SP skew and Noise-to-Signal Ratio (NSR)—parameters highlighted on the green trace of Figure \ref{F8}a and summarized in Table 2 in Supporting Information. Consequently, the fitted \textit{soft-bound} model does not account for inter-device variability. Finally, the device presets were incorporated into a 
3-layer Fully Connected (3-FC) NN for MNIST digit classification \cite{MNIST}.

Results depicted in Figure \ref{F8}b show the decrease of the obtained training accuracy when the SP shifts away from the $G$-window center, leading to non-convergent behavior for the extreme case of highly shifted SP (green boxes in Figure \ref{F8}b). Highest accuracy and fastest convergence are obtained for strongly-centered SP, which is obtained when employing $V_{\rm P+}=+1.47$ V and $V_{\rm P-}=-1.45$ V. This pulse-amplitude setup leads to a limited lower bound $G_{\rm min}$, and therefore a smaller $G$-window, due to weaker conductance changes during the depression phase, see the yellow trace in Figure \ref{F8}a. The smaller steps allow for accessing more conductance states within that conductance range, and therefore, preserving the number of available states for training. \\


The findings outlined in this section show that pulse amplitude tuning can compensate for the relative abruptness between bidirectional switching phases, thereby improving the symmetry of the switching characteristics. Although the 'aihwkit' training simulation does not explicitly account for pulse amplitudes—considering only their polarity and effect on conductance change—achieving high symmetry can lead to a fast training convergence with TTv1 and near-floating-point accuracy with minimal digital overhead. Enhancing the electrical response, through pulse scheme optimization, provides an alternative to the more complex algorithms, offering increased flexibility in scenarios where higher computational costs may not be achievable.

\section{Conclusion}

In this work, an advanced physics-based compact model for analog filamentary CMO/HfO$_{\rm x}$ ReRAM is presented. It accurately captures analog resistive switching phenomena in the ReRAM memory cell over a wide range of operating conditions, reproducing multiple experimental characteristics with negligible error thanks to the physical framework of the model core. Such input conditions-independent accuracy makes this compact model suitable for predictive simulations in circuit-level design environments. \\

Compared to the previous version \cite{ESSDERC24_Galetta}, the current model incorporates the following key enhancements. First, it accounts for parasitic resistive elements external to the ReRAM cell, which influence the measured switching characteristics. Second, the ion migration dynamics have been corrected to prevent any unphysical concentration levels of oxygen-related defects, ensuring realistic switching profiles. Third, the model introduces dynamically evolving physical parameters that govern hopping conduction, based on the real-time configuration of defects—an essential feature to accurately capture the analog behavior of the device. These enhancements enable the model to replicate a wide spectrum of experimental behaviors, including quasi-static $I$-$V$ curves, single-pulse SET kinetics, and analog bidirectional conductance modulation. \\

In the quasi-static domain, simulations of the ion drift mean velocity profile provide insights into the average defect migration displacements during switching transitions. Single-pulse programming simulations are used to study the switching kinetics of CMO/HfO$_{\rm x}$-based ReRAM and to derive insights into thermal trade-offs when using short, low-energy pulses with high programming voltages. Additionally, the compact model supports realistic switching non-idealities by integrating a variability component for stochastic TAT mechanisms.

A key highlight of the study is the analysis of the symmetry point within the $G$-window of analog bidirectional accumulative conductance responses. The model reveals that equilibrium memory states arise from a balance of opposing forces driving defect migration, which can be tuned via input pulse modulation. The physical origin of this stabilization—and its manifestation in both quasi-static $I$-$V$ characteristics and pulse responses—is further explored through the dynamic route map approach, which enables visualization and mapping of equilibrium memory states across different input domains.\\

The model’s predictive capabilities are employed in neuromorphic computing contexts. Different pulse schemes are used to achieve nearly perfect switching symmetry, characterized by a symmetry point skew around 50\%. The impact of symmetry point tuning—achieved by adjusting the voltage pulse amplitudes—was validated for on-chip training, when employing the earliest version of the Tiki-Taka algorithm, with minimal digital overhead. Simulation results demonstrated that when pulse-driven switching symmetry is obtained, near-floating-point performance is achieved in a 3-layer Fully Connected Neural Network for MNIST digit classification. These results underscore the importance of input pulse engineering for analog ReRAM-based RPUs. \\

Future work will focus on integrating the proposed compact model into standard IC design tools, such as SPICE-based simulators, which require differentiable device descriptions. Owing to its inherent compatibility, the model presented in this study lays the groundwork for circuit-level simulations including CMO/HfO$_{\rm x}$-based ReRAM technology, thereby supporting its application in both analog computing and digital non-volatile memory architectures.

\section{Experimental Section}

\textit{Electrical Characterization}: \\
The quasi-static $I$-$V$ sweep characterization of the CMO/HfO$_{\rm x}$-based ReRAM device is performed with an Agilent B1500A Semiconductor Device Parameter Analyzer. A voltage signal is applied to the device's top electrode, while the bottom electrode is biased to the earth ground. The quasi-static signal consists of a staircase ramp with voltage steps of 10 mV, each long 100 ms/step (measurement time), therefore having a global Sweep Rate (SR) of 0.1 Vs$^{-1}$. Two source measurement units are used, each of which simultaneously biases the ReRAM device and measures the current flowing into each terminal (top or bottom electrode). No external resistors are present in series with the ReRAM device during quasi-static characterizations. \\
Pulse responses (single-pulse programming and bidirectional accumulative conductance response) characterizations are conducted using a 16-bit 400 MS/s NI PXIe-5451 arbitrary waveform generator to deliver the generated pulses to the device's top electrode, and a NI PXIe-5164 oscilloscope with 1 GS/s sampling rate and 400 MHz bandwidth to measure the current signal. The current signal is measured through the 50 $\Omega$ oscilloscope internal resistor, in series with the ReRAM device bottom electrode. To cancel out any potential measurement offset, experimental read pulses, according to the electrical pulse schemes reported in the manuscript, are divided into two alternating positive and negative pulses with amplitudes of $\pm$200 mV.

\textit{Fabrication process}: \\
The fabrication process of the 200 nm $\times$ 200 nm ReRAM cell consisting of a TiN/TaO$_{\rm x}$/HfO$_{\rm x}$/TiN material stack (in order from the top electrode towards the bottom electrode) is described in our previous works \cite{Stecconi2022}.

\medskip
\textbf{Supporting Information} \par 
Supporting Information is available from the Wiley Online Library or from the author.

\medskip
\textbf{Acknowledgements} \par
The authors acknowledge the Binnig and Rohrer Nanotechnology Center (BRNC) at IBM Research Europe - Zürich. The authors would like to thank Linda Rudin for proofreading the manuscript. \\
This work is funded by the European Union and Swiss state secretariat SERI within the PHASTRAC (grantID: 101092096) project, by SNSF ALMOND (grantID: 198612), and by the Federal Ministry of Education and Research (BMBF, Germany) within the NEUROTEC (grantID: 16ME0398K) project.

\textbf{Author contributions} \par
Conceptualization: M. G., D. F. F. and V. B.; electrical characterization: M. G., D. F. F and F. H.; device fabrication: T. S.; compact model simulations: M. G.; training simulations: V. C.; result interpretation: M. G., D. F. F., V. C., W. C., S. M., A. L. P., B. J. O. and V. B.; supervision: V. B. and B. J. O.; manuscript writing: M. G. and V. C.; data curation: M. G., D. F. F., V. C. and V. B.; manuscript review and editing: all authors; funding acquisition: V. B. and B. J. O.

\medskip
\textbf{Conflict of Interest} \par
The authors declare no competing interests.

\medskip

\bibliographystyle{MSP}
\bibliography{my_bibliography.bib}


\begin{figure}[H]
\textbf{Table of Contents}\\
\medskip
  \includegraphics[width=13cm]{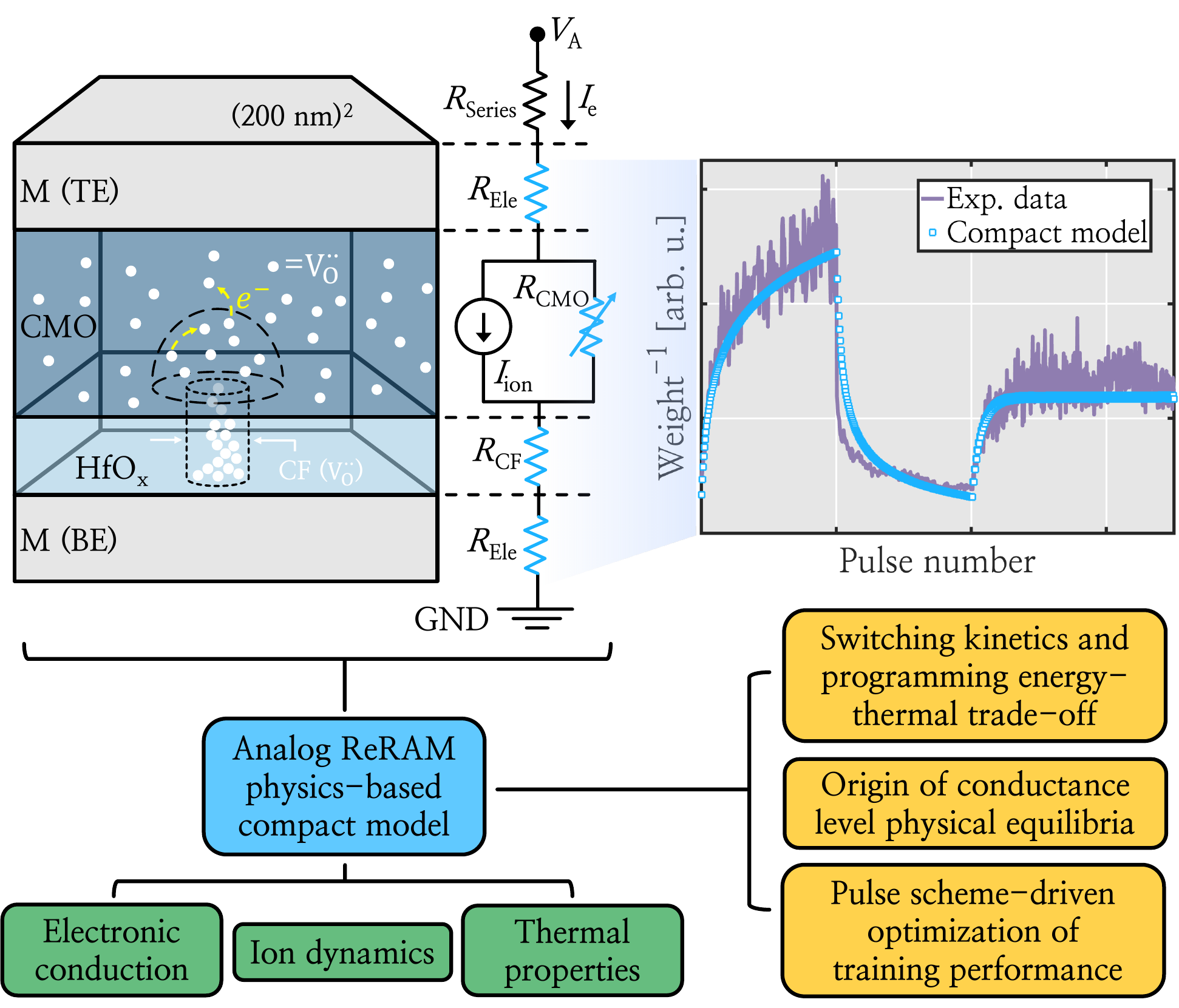}
  \medskip
  \caption*{ A physics-based compact model for Conductive-Metal-Oxide/HfO$_{\rm x}$ ReRAM, accounting for ion dynamics, electronic conduction, and thermal effects, is presented. Accurate and versatile simulations of analog non-volatile conductance modulation and memory state stabilization enable reliable circuit-level studies, advancing the optimization of neuromorphic and memory systems, driven by device and material physics understanding.}
\end{figure}

\end{document}